\documentclass[pre,preprint,showpacs]{revtex4}
\usepackage{amssymb}
\usepackage[latin1]{inputenc}
\usepackage{graphicx}
\usepackage{bm}

\begin{document}

\title{Thermal segregation of intruders in the Fourier state of a granular gas}
\author{J. Javier Brey and Nagi Khalil}
\affiliation{FÝsica Te¾rica, Universidad de Sevilla, Apartado de Correos 1065, E-41080
Sevilla, Spain}
\author{James W. Dufty}
\affiliation{Department of Physics, University of Florida, Gainesville, FL 32611, USA}

\begin{abstract}
A low density binary mixture of granular gases is considered within the
Boltzmann kinetic theory. One component, the intruders, is taken to
be dilute with respect to the other, and thermal segregation of the
two species is described for a special exact solution to the
Boltzmann equation. This solution has a macroscopic hydrodynamic
representation with a constant temperature gradient and is referred
to as the Fourier state. The thermal diffusion factor characterizing
conditions for segregation is calculated without the usual
restriction to Navier-Stokes hydrodynamics. Molecular dynamics
simulations are reported for comparison with the results for this
idealized Fourier state.

\end{abstract}

\date{\today }
\maketitle

\section{Introduction}
Segregation among species in a granular mixture is of significant
practical importance, and its description, both qualitative and quantitative,
a challenging problem \cite{Ku04}. Many physical
systems and state conditions are of interest, involving a range of
mechanisms responsible for segregation across a wide range of
control parameters. Here, attention is focused on a simple system of
gas mixtures in a temperature gradient, being the latter the only
mechanism available to induce segregation. Furthermore, each species
is assumed to be composed of smooth, inelastic hard spheres at low
density, so the inelastic Boltzmann kinetic equation applies
\cite{GyS95}. One component (the impurities) is taken to be dilute
with respect to the other (the host). In this limit the host gas is
not affected by the impurities, and is described by its own independent
Boltzmann equation.

The macroscopic balance equations, or hydrodynamic equations,
obtained from this kinetic theory description have special exact
solutions: the host gas
has zero flow velocity, a constant temperature gradient and a constant pressure \cite%
{BCMyR01,BKyR09}; the impurity gas has zero flow velocity, a
temperature proportional to the host temperature, and a pressure
proportional to a power of that temperature \cite{BKyD11}. Since the
heat flux of the host component can be expressed as proportional to the
temperature gradient, i.e. by Fourier's law, this state is referred to
here as the Fourier state. The coefficients in this hydrodynamic
description are defined in terms of\ coupled, nonlinear integral
equations obtained from kinetic theory. In this way, an exact
description of a mixture in a temperature gradient is obtained as
the basis for exploration of segregation as a function of the
temperature gradient and differences in mechanical properties of the
two species. No explicit limitations on the magnitude of the
temperature gradient are assumed in this analysis and all higher
degree derivatives are exactly zero. This complements the extensive
studies of thermal segregation based on applications of kinetic
theory at the Navier-Stokes level \cite{GyR09,G11}.

Due to the symmetry of the problem, all spatial variation of
properties occurs along the temperature gradient, taken to be the
$x$ axis. The segregation of impurity particles relative to the host
gas is described by the variation of the composition
$n_{0}(x)/n(x)$, where $n_{0}(x)$ and $n(x)$ are the impurity and
host densities, respectively. In the absence of the temperature
gradient, these densities are uniform and there is no segregation.
For a finite temperature gradient, a convenient measure of
thermal segregation is given in terms of the thermal diffusion factor $%
\Lambda $ defined by%
\begin{equation}
\frac{d}{dx}\ln \frac{n_{0}(x)}{n(x)}=-\Lambda \frac{d\ln
T(x)}{dx}\,. \label{1.1}
\end{equation}%
For the special hydrodynamic state constructed here this
dimensionless factor is independent of $x$ and therefore a global
property of the system. For more general states, $\Lambda $ is a
local function and segregation properties can vary throughout the
system. Here, it can depend only on a dimensionless form of the
temperature gradient as well as ratios of the mechanical properties
of each species (size, mass, degree of collisional inelasticity). It
can be positive or negative, implying that the impurity
concentration increases against or along the temperature gradient,
respectively (a thermal analogue of the Brazil nut and reverse
Brazil nut effects for gravitational segregation
\cite{DRyC93,HKyL01,JyY02,BRyM05}). Of course, $\Lambda $ should
vanish for mechanically identical host and impurity particles.

The expression for $\Lambda $ depends on coefficients defining the
hydrodynamic fields which in turn are defined in terms of solutions
to nonlinear integral equations following from the underlying kinetic theory.
The kinetic theory and associated hydrodynamics for the Fourier
state has been described elsewhere \cite{BKyD11}, so only a brief
summary is presented in the next section. An analytic
solution to the integral equations can be obtained as described in \cite%
{BKyD11}, and the results are extended here to arbitrary dimension
$d$ in the Appendix. More generally, for arbitrary degree of
inelasticity the equations are solved by a truncated Sonine
polynomial expansion \cite{BKyR09,BKyD11}. The results in both cases
for $\Lambda $ and the violation of energy equipartition  measured by means of
the parameter $\gamma
 \equiv T_{0}/T$, are illustrated for a wide range of degrees of
inelasticity (granular and non-equilibrium systems violate the
equipartition property of equilibrium states and hence the steady
temperatures of the two systems are different in general). Next,
event driven molecular dynamics simulations (MD) are described for
$200$ inelastic hard disks ($d=2$) in a rectangular box with thermal
walls to generate a temperature gradient. The Fourier state spatial
dependence of the hydrodynamic fields for both species is confirmed
in the bulk of the system, away from boundary layers near the walls.
Comparison of the \ MD and Fourier state results for the temperature ratio $%
\gamma $ as a function of dissipation, mass ratio, and size ratio
shows good agreement only for weak to moderate dissipation, or small
mechanical differences, but only qualitative agreement is found otherwise.
This poor accuracy is interpreted as a limitation of the Sonine
approximation for this property.
However, MD and Fourier state results for the thermal diffusion factor $%
\Lambda $ show much better agreement over the entire parameter
space. These results are discussed further in the last section.

\section{The Fourier state for the host gas and the impurities}
\label{s2}Consider a system of $N$ smooth inelastic hard spheres
($d=3$) or disks ($d=2$) of mass $m$ and diameter $\sigma $.
Inelasticity of collisions is characterized by a constant, velocity
independent, coefficient of normal restitution $\alpha $, defined in
the interval $0<\alpha \leq 1$. Then, when two particles, $i$ and
$j$, with velocities ${\bm v}_{i}$ and ${\bm v}_{j}$ collide, the
velocities are instantaneously modified to new values given by
\begin{eqnarray}
{\bm v}_{1}^{\prime } &=&{\bm v}_{1}-\frac{1+\alpha }{2}({\bm
g}\cdot
\widehat{\bm\sigma })\widehat{\bm\sigma },  \nonumber \\
{\bm v}_{2}^{\prime } &=&{\bm v}_{1}+\frac{1+\alpha }{2}({\bm
g}\cdot \widehat{\bm\sigma })\widehat{\bm\sigma },  \label{2.1}
\end{eqnarray}%
where ${\bm g}\equiv {\bm v}_{1}-{\bm v}_{2}$ and
$\widehat{\bm\sigma }$ is the unit vector joining the centers of the
two particles at contact. The system is supposed to be very dilute,
so that the one-particle distribution
function for position ${\bm r}$ and velocity ${\bm v}$ at time t, $f({\bm r},%
{\bm v},t)$, obeys the inelastic Boltzmann equation \cite{GyS95}. In \cite%
{BCMyR01,BKyR09} a special solution of this equation was proposed,
and the state associated with it was called the Fourier state. It is
a time-independent distribution function with gradients only in one
direction and having a scaling form in terms of the hydrodynamic
fields,
\begin{equation}
f(x,{\bm v})=n(x)\left[ \frac{m}{2 T(x)}\right] ^{d/2}\phi ({\bm c}),\quad {%
\bm c}\equiv \left[ \frac{m}{2T(x)}\right] ^{1/2}{\bm v}.
\label{2.2}
\end{equation}%
In the above expressions, $T$ is the granular temperature and $n$ is
the number of particles density. Both are defined from the
one-particle distribution function in the usual way, although with
the Boltzmann constant set equal to unity. It is verified that the
state defined by the above distribution has a uniform hydrodynamic
pressure $p\equiv n(x)T(x)$, a heat flux $q_{x}$ proportional to $T(x)^{1/2}$, and it
exhibits a linear temperature profile,
\begin{equation}
\frac{dT(x)}{dx}=Bp\sigma ^{d-1},  \label{2.3}
\end{equation}%
with the dimensionless constant $B$ given by a functional of $\phi $ \cite%
{BCMyR01,BKyR09}. This functional, as well as the function $\phi $,
was approximately determined by solving the Boltzmann equation
using a representation of the distribution function $\phi $ as a
truncated Sonine polynomial expansion, and keeping only up to
bilinear terms in the coefficients \cite{BKyR09}. The results were
shown to be in good agreement
with molecular dynamics simulation data, at least for weak inelasticity ($%
0.9\leq \alpha <1$).

Now suppose that $M$ additional hard spheres or disks of mass
$m_{0}$ and
diameter $\sigma _{0}$ are added to the host gas in the Fourier state. For $%
M\ll N$, the effect of these \textquotedblleft
impurity\textquotedblright\ particles or \textquotedblleft
intruders\textquotedblright\ on the host gas distribution function
is negligible, so that the
results described above remain valid. Moreover, the dynamics of the
impurities is determined by their collisions with the host gas
particles, while collisions between intruders can be neglected. When
an intruder with velocity ${\bm v}_{0}$ collides with a gas particle
of velocity ${\bm v}_{1}$, the velocities are instantaneously
changed into
\begin{eqnarray}
{\bm v}_{0}^{\prime } &=&{\bm v}_{0}-\frac{m(1+\alpha _{0})}{m+m_{0}}%
\,\left( {\bm g}_{0}\cdot \widehat{\bm\sigma }\right)
\widehat{\bm\sigma },
\nonumber \\
{\bm v}_{1}^{\prime } &=&{\bm v}_{1}+\frac{m_{0}(1+\alpha _{0})}{m+m_{0}}%
\,\left( {\bm g}_{0}\cdot \widehat{\bm\sigma }\right)
\widehat{\bm\sigma }, \label{2.3a}
\end{eqnarray}%
where ${\bm g}_{0}\equiv {\bm v}_{0}-{\bm v}_{1}$ and $\alpha _{0}$
is the coefficient of normal restitution for collisions between an
intruder and a host gas particle. It is also defined in the interval
$0<\alpha _{0}\leq 1$. In the tracer limit being considered, the
one-particle distribution function for the additional particles,
$F({\bm r},{\bm v}_{0},t)$, obeys the inelastic Boltzmann-Lorentz
equation \cite{RydL77,BKyD11}. A solution of this equation similar
to (\ref{2.2}) is considered,
\begin{equation}
F(x,{\bm v}_{0})=n_{0}(x)\left[ \frac{m_{0}}{2\gamma T(x)}\right]
^{d/2}\Phi
({\bm c}_{0}),\quad {\bm c}_{0}\equiv \left[ \frac{m_{0}}{2\gamma T(x)}%
\right] ^{1/2}{\bm v}_{0}.  \label{2.5}
\end{equation}%
Here $n_{0}$ is the number density of intruders and the function
$\Phi $ has been chosen such that the average velocity vanishes and
the granular
temperature of the impurities, defined from the second velocity moment of $F$%
, is $T_{0}(x)=\gamma T(x)$. The parameter $\gamma $, measuring the
deviation from energy equipartition, must be identified (similarly
to the function $\Phi $) by requiring Eq. (\ref{2.5}) to be a
solution of the Boltzmann-Lorentz kinetic equation. A necessary
condition for it is that \cite{BKyD11}
\begin{equation}
\frac{d\ln n_{0}(x)}{dx}=Cn(x)\overline{\sigma }^{d-1},  \label{2.6}
\end{equation}%
with $\overline{\sigma }\equiv (\sigma +\sigma _{0})/2$ and $C$
being a constant. The distribution function $\Phi $ and the constant
$C$ have also
been evaluated by using a truncated Sonine representation for $\phi $ and $%
\Phi $ \cite{BKyD11}. The distribution function $F $ is
\textquotedblleft normal\textquotedblright\ in the context of
kinetic theory, i.e. it depends on position only through the local
densities and the temperature of the system.

As already mentioned in the Introduction, the main focus in this
paper is on segregation, i.e. the demixing of the granular mixture, by
thermal diffusion. The amount of segregation can be measured by the
thermal diffusion factor $\Lambda $ defined by (\ref{1.1}). The
value $\Lambda =0$ indicates that no segregation induced by the
temperature gradient occurs, in the sense that the density ratio
does not depend on position. On the other hand, when $\Lambda >0$
the impurity concentration increases against the temperature
gradient, while for $\Lambda <0$ the impurity concentration
increases as the temperature increases. By means of Eqs.\ (\ref{2.3}) and (%
\ref{2.5}) the thermal diffusion factor can be expressed in the
present case as
\begin{equation}
\Lambda =-\left( 1+\frac{\overline{\sigma }^{d-1}C}{\sigma
^{d-1}B}\right) . \label{2.8}
\end{equation}

For arbitrary values of the restitution coefficients, the equations
determining the coefficients of the truncated Sonine expansions of
the distribution functions are rather involved (even in the bilinear
approximation used in refs. \cite{BKyR09} and \cite{BKyD11}) and
they must be solved numerically. Nevertheless, in the limit of small
inelasticity for the host gas, i.e. $1 -\alpha \ll 1$, it is
possible to obtain explicit expressions for the coefficients. The
results are summarized in the Appendix. In Figs. \ref{fig1} and
\ref{fig2}, the asymptotic expressions for $\Lambda$ and $\gamma$
for small inelasticity of the host gas, as given in the
Appendix, are compared with the numerical solution of the kinetic
equations. The comparison is carried out as a function of the
coefficient of inelasticity $\alpha$ for several values of
$\alpha_{0}$, as indicated in
the figures. The constant values of the other parameters are $d=2$, $m_{0}=m$%
, and $\sigma_{0}= \sigma$. It is important to realize that all the
expressions being compared have been obtained in the Sonine
approximation and neglecting products of three or more of the
coefficients appearing in the truncated expansion of the
distribution function. Consequently, nothing can be concluded about
the validity of the truncated Sonine approximation from those figures.
It is seen that the accuracy of the
asymptotic expression for the temperature ratio extends up to
smaller values of the coefficient $\alpha$ than the corresponding expression
 for the thermal
diffusion factor. Actually, the latter does not depend on $\alpha$ in the
considered limit. This behavior was to be expected since, as pointed out in
the Appendix, the former has been evaluated to order $1-\alpha$ and
the latter to order $(1-\alpha)^{1/2}$. Similar results are obtained
for inelastic hard spheres ($d=3$).

\begin{figure}
%[tbp]
\includegraphics[scale=0.6,angle=-90]{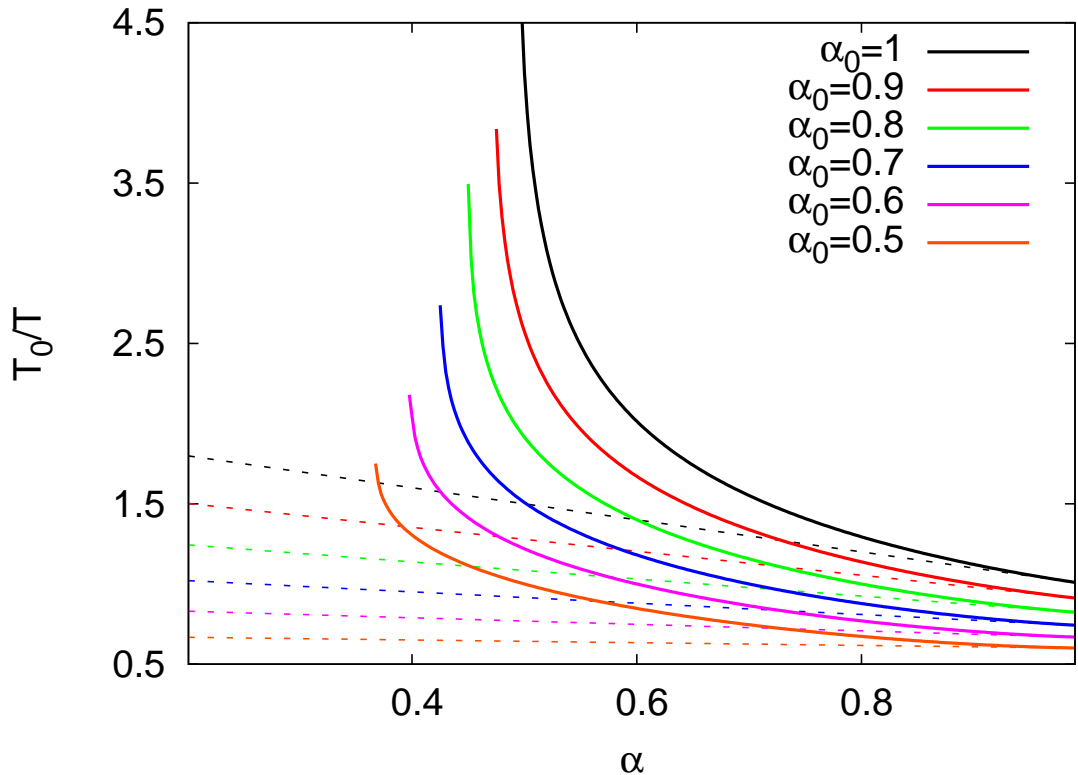}
\caption{(Color on line) Temperature of the impurities $T_{0}$
divided by the temperature of the host gas $T$ as a function of the
coefficient of normal restitution of the gas particles
$\protect\alpha$, for several values of the restitution coefficient
for collisions between the gas particles and the impurities,
$\protect\alpha_{0}$, as indicated in the insert (the lower the
curve the smaller $\protect\alpha_{0}$). In all cases, $d=2$,
$m_{0}=m$, and $\protect\sigma =\protect\sigma
_{0}=\overline{\protect\sigma } $. The solid lines have been
obtained by solving numerically the Boltzmann equation in the Sonine
approximation, while the dashed ones are the expressions in the
almost elastic host gas limit given in the Appendix.} \label{fig1}
\end{figure}

\begin{figure}%[tbp]
\includegraphics[scale=0.6,angle=-90]{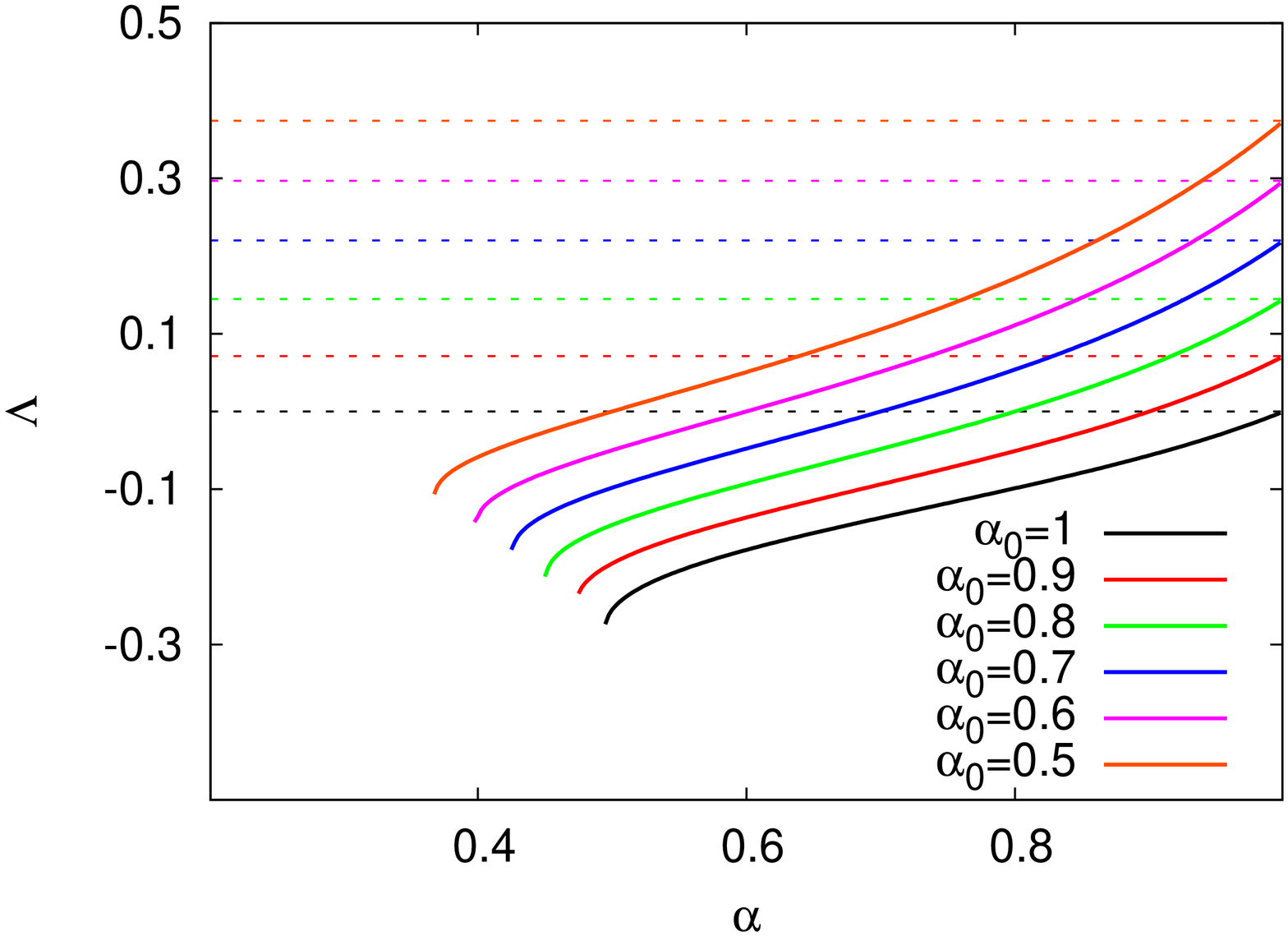}
\caption{(Color on line) The dimensionless thermal diffusion factor
$\Lambda$
as a function of the coefficient of normal restitution of the gas particles $%
\protect\alpha$, for several values of the restitution coefficient
for
collisions between the gas particles and the impurities, $\protect\alpha_{0}$%
, as indicated in the insert (the lower the curve the larger $\protect\alpha%
_{0}$). In all cases, $d=2$, $m_{0}=m$, and $\protect\sigma =\protect\sigma %
_{0}=\overline{\protect\sigma } $. The solid lines have been
obtained by solving numerically the Boltzmann equation in the Sonine
approximation, while the dashed ones are the expressions in the
almost elastic host gas limit given in the Appendix.} \label{fig2}
\end{figure}

\section{Molecular Dynamics simulations}
\label{s3} Event driven molecular dynamics simulations of a system
of inelastic hard disks ($d=2$) have been performed to check the
accuracy of the above theoretical predictions. In all the
simulations to be reported, the number of particles is $N=200$ and
they are enclosed in a rectangular cell of sides
$L_{x}=4L_{y}=200\sigma $. Moreover, only a single impurity was
considered, i.e. $M=1$. The simulations started with all the
particles, including the impurity, uniformly distributed on a square
lattice and with a Gaussian velocity distribution.

\subsection{The Fourier state}

To generate the Fourier state, the same procedure as in ref.
\cite{BKyR09} was employed. To inject energy into the system and
induce the temperature gradient, two thermal walls
\cite{Ce69,DyvB97} located at $x=0$ and $x=L_{x}$ were considered,
while periodic boundary conditions were used in the $y$ direction.
The temperatures of the walls were chosen accordingly with the
theoretical prediction for the Fourier state \cite{BKyR09}. For the
cases being reported, it was found that the system reached, after a
transient period, a steady state with gradients only in the
$x$-direction and no macroscopic flow. This is consistent with the
value chosen for the aspect ratio $L_{x}/L_{y}$, which is outside
the region in which the transversal hydrodynamic instability
exhibited by the state we are considering shows up
\cite{LMyS02,BRMyG02} for the parameters used in the simulations.

\begin{figure}[tbp]
\includegraphics[scale=0.4,angle=-90]{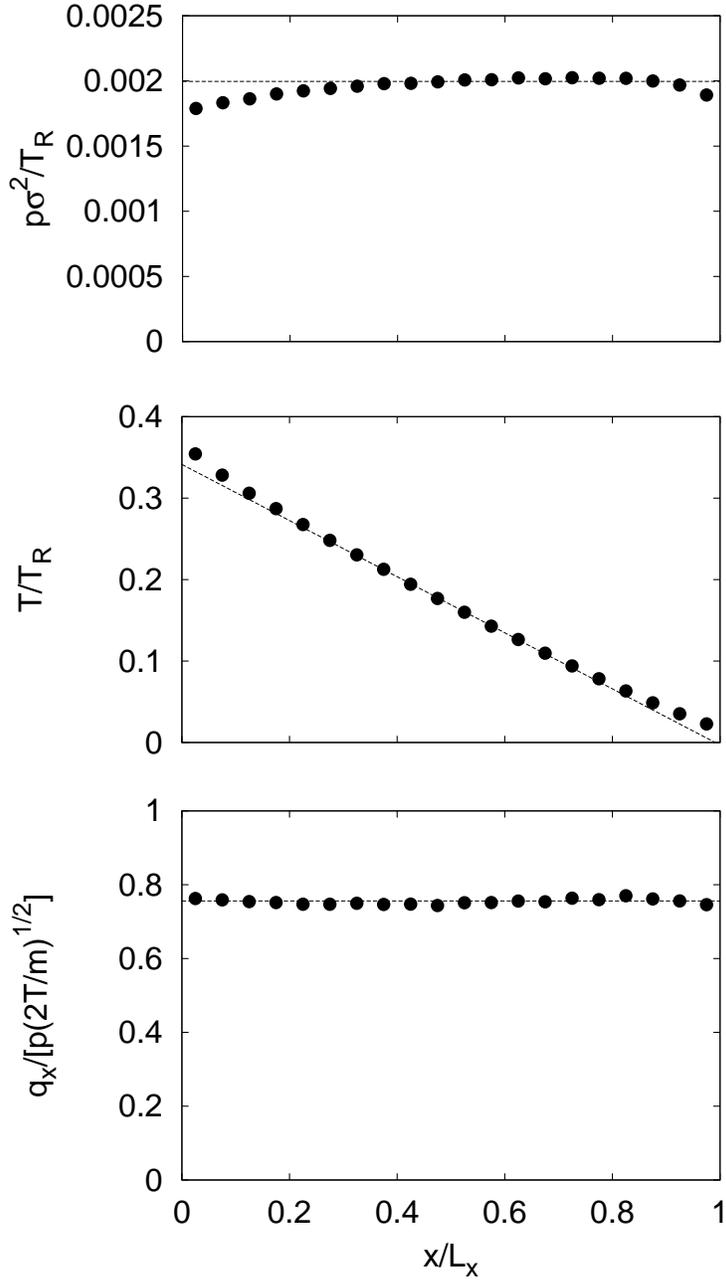}
\caption{Steady dimensionless pressure, temperature
and heat flux profiles for a system of hard disks with
$\protect\alpha=0.9$. The mechanical properties of the impurity are
the same as those of the host gas particles. The temperature and
pressure have been scaled with some arbitrary reference value,
$T_{R}$, actually the initial temperature of the system. The symbols
are simulation data, the dashed straight lines in the pressure and
heat profiles are guides for the eye, and the dashed line in the
temperature profile is a linear fit of the data in the bulk of the
system.} \label{fig3}
\end{figure}

Figure \ref{fig3} shows the pressure, temperature, and heat flux
profiles measured in a system with $\alpha =0.9$. The finite system
generated by MD has boundary layers near $x=0$ and $x=L_{x}$, where a
hydrodynamic description does not hold. Outside those layers a bulk
region in which the pressure and the scaled heat flux are uniform while the
temperature is linear is clearly identified. This is the region in
which the theoretical predictions are expected to hold. For $\alpha
>0.9$ a similar behavior is observed, with the size of the bulk
region increasing as $\alpha $ increases.

Once the system has reached the Fourier state, several properties of
the gas and the impurity have been measured. The system has been
divided in 20 slices parallel to the $y$ axis of the same width.
Moreover, the results that will be shown have been averaged in time,
and also over $500$ trajectories of the system. The emphasis has been
put on the temperature and density profiles of both the gas and the
impurity, and on the values of the parameters $\gamma $ and $\Lambda
$ obtained from them. In order to get a systematic information of
the role played by the several properties of the impurity ($\alpha
_{0},m_{0},\sigma _{0}$), only one of these properties is chosen in
each case to differ from the gas host particles. Then, for
instance, in those cases in which $\alpha \neq \alpha _{0}$, the values $%
m_{0}=m$ and $\sigma _{0}=\sigma $ have been employed.

\begin{figure}[tbp]
\includegraphics[scale=0.5,angle=-90]{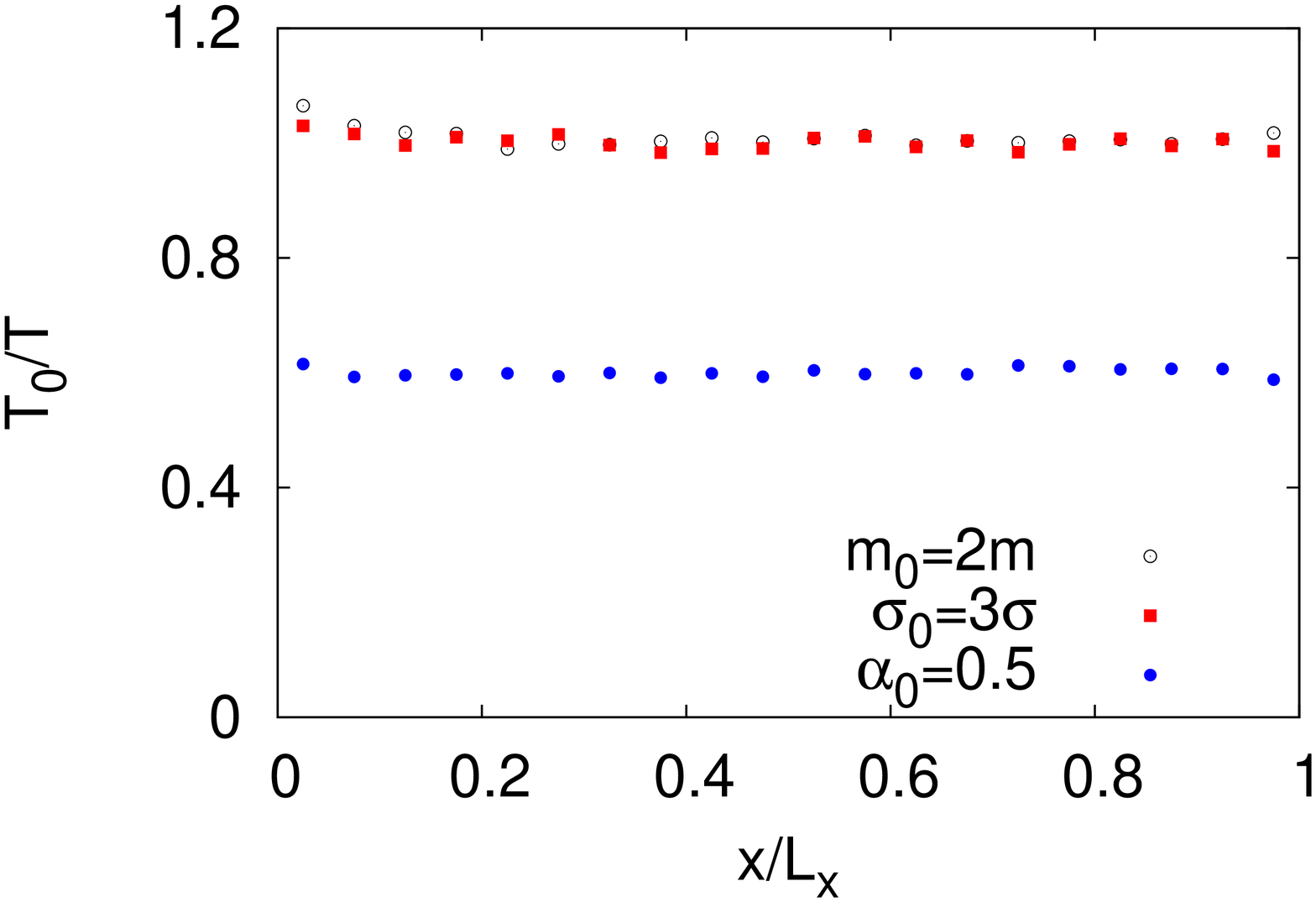}
\caption{(Color on line) Profile along the $x$ direction of the
ratio between the temperature of the intruder $T_{0}$ and the
temperature of the system $T$ for a system with $\protect\alpha
=0.99$.} \label{fig4}
\end{figure}

The quantity $\gamma \equiv T_{0}/T$ has been computed by
identifying the bulk region of the system in which the temperature
ratio is homogeneous. An example for $\alpha =0.99$ is given in
Fig.\ \ref{fig4}. The different data sets correspond to vary one of
the parameters of the impurity as indicated in the insert. In each
of the cases, all the others parameters are the same for the host
gas particles and for the impurity. It is seen that the ratio is
uniform over most of the system. The deviation from unity is small
for the differences in mass and size ratios considered since the
system is almost elastic in all collisions. However, when there is a
large difference between the two coefficients of normal restitution
a rather strong violation of energy equipartition shows up. This is
a general property of granular mixtures
\cite{DRyC93,FyM02,GyD99,ByR11}. The values of $\gamma $ reported in
the following have been obtained by averaging the temperature ratio
in the region in which it is roughly uniform.

\begin{figure}[tbp]
\includegraphics[scale=0.5,angle=-90]{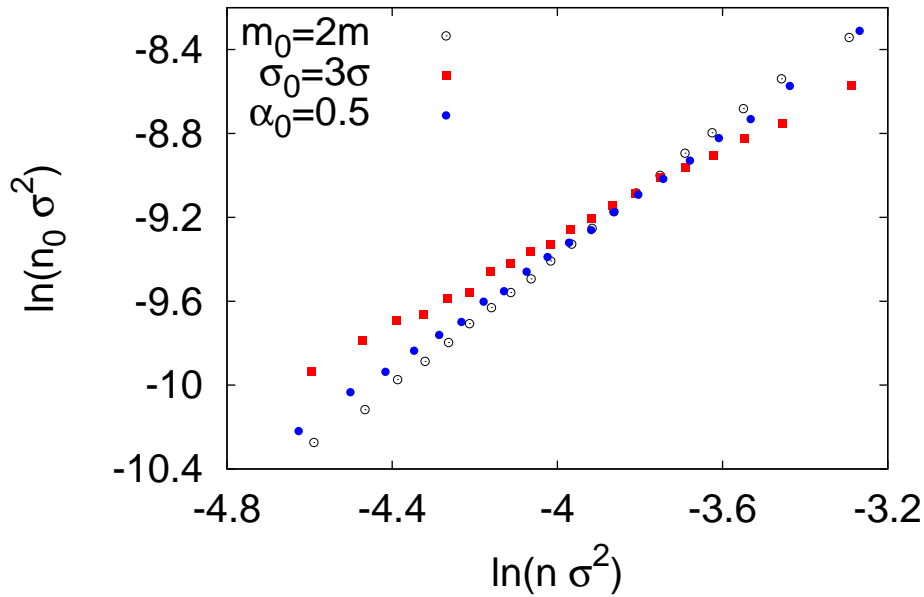}
\caption{(Color on line) Logarithm of the (dimensionless) intruder density as a function of the
logarithm of the host gas (dimensionless) density, for a system with $\protect\alpha
=0.99$.} \label{fig5}
\end{figure}

The coefficient $\Lambda $ could be measured, in principle, by using
its definition in Eq. (\ref{1.1}). Nevertheless, the measurement of
temperatures from the simulation data introduces more uncertainties
than the measurement of densities. Taking into account that in the
Fourier state the pressure is uniform, Eq. (\ref{1.1}) can be
transformed into
\begin{equation}
\Lambda =\frac{d\ln n_{0}}{d\ln n}-1.  \label{3.1}
\end{equation}%
This is the expression actually used to compute $\Lambda $, i.e. it
is
obtained from the slope of linear profile of $\ln n_{0}$ as a function of $%
\ln n$. An example is provided in Fig. \ref{fig5}. The three sets of data
correspond to the intruder differing from the host gas particles in
the mechanical property indicated in the insert.

\subsection{Comparisons for $\protect\gamma $ and $\Lambda $}

\begin{figure}[tbp]
\includegraphics[scale=0.5,angle=-90]{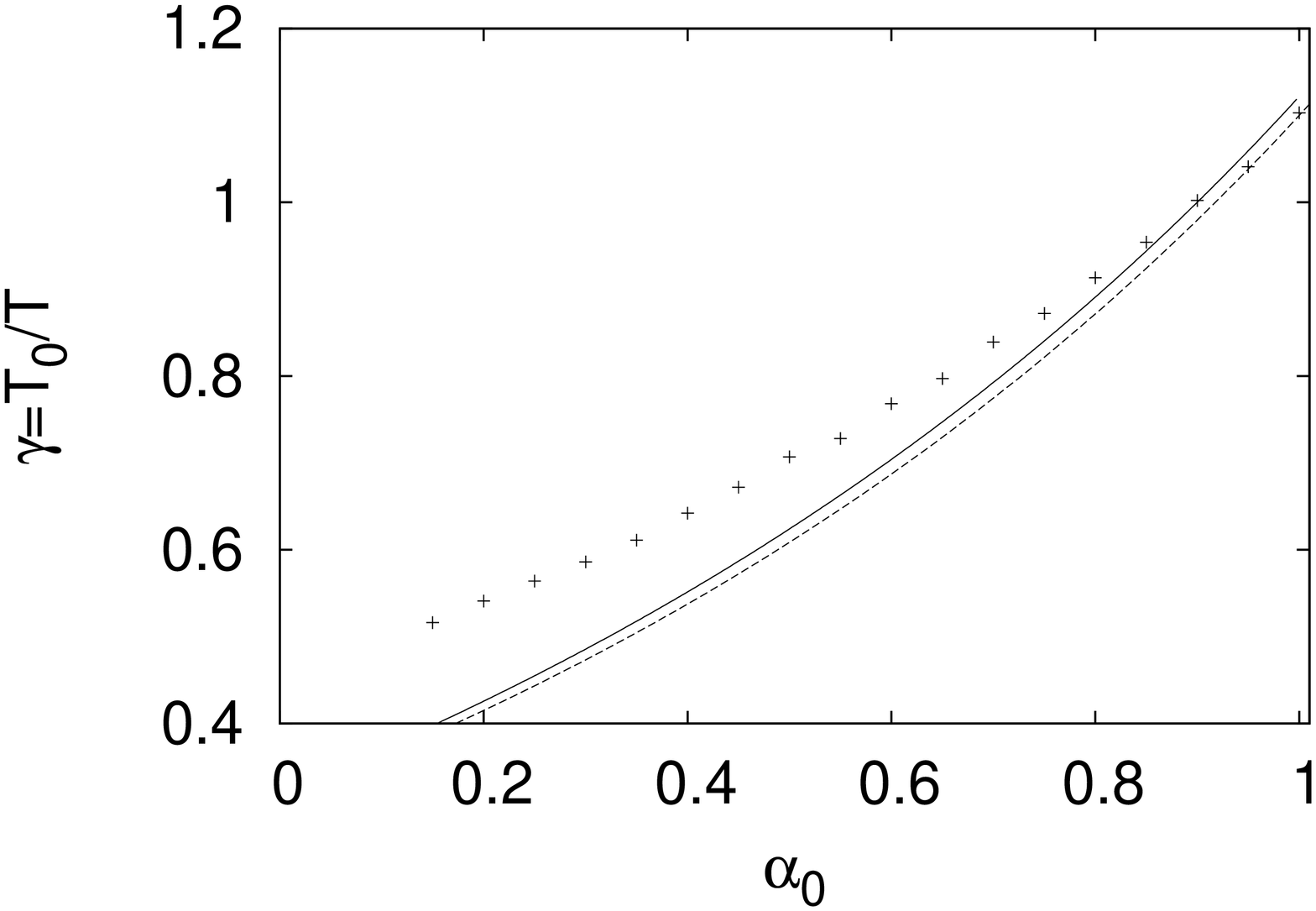}
\caption{Ratio between the impurity temperature $T_{0}$ and the host
gas temperature $T$ in the bulk of the system, once the Fourier
state has been reached, as a function of the coefficient of normal
restitution for the collisions between the intruder and the host gas
particles. The coefficient of normal restitution for the gas is
$\protect\alpha=0.9$. The mass and diameter of the intruder are the
same as those of the host particles. The symbols are MD
simulation results, the solid line the theoretical
prediction from the Boltzmann equation using a truncated Sonine
expansion, and the dashed line the weak inelasticity host gas limit
given in the Appendix and discussed in the text.} \label{fig6}
\end{figure}

\begin{figure}[tbp]
\includegraphics[scale=0.5,angle=-90]{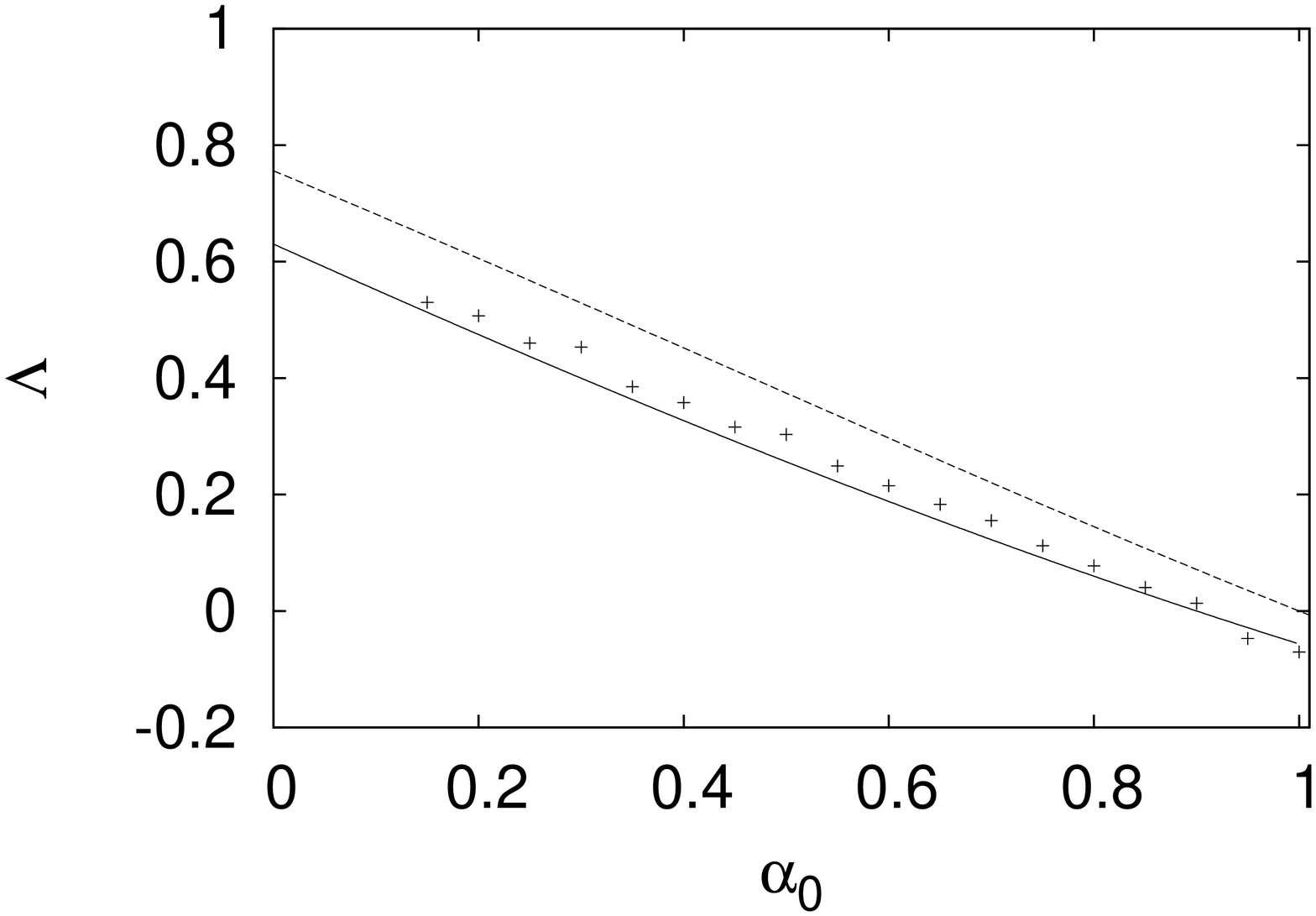}
\caption{The dimensionless thermal diffusion factor $\Lambda$ for
the same system as in Fig. {\protect\ref{fig6}}.} \label{fig7}
\end{figure}

In the following, results will be restricted to the case
$\alpha=0.9$. Figures \ref{fig6} and \ref{fig7} show the behavior of
$\gamma$ as a function of $\alpha_{0}$ ($m=m_{0}$ and
$\sigma=\sigma_{0}$) and of $m_{0}/m$ ($\alpha= \alpha_{0}$ and
$\sigma = \sigma_{0}$), respectively. A good agreement between the
theoretical predications and the simulation data is observed,
although quantitative discrepancies in $\gamma$ appear for small
values of $\alpha_{0}$, i.e. very inelastic intruder. Figure
\ref{fig7} demonstrates the existence of segregation. The thermal
diffusion coefficient is positive when $\alpha_{0} < 0.9$, i.e. the
impurity concentration is higher at the colder part of the system.
For larger values of $\alpha_{0}$, segregation occurs in the
opposite direction; the impurity concentration is
higher in the hotter part of the system. Consistently, the value of $%
\alpha_{0}$ for which the direction of the segregation effect changes is
also the value at which the temperature of the impurity equals the
temperature of the host gas, since at this point the intruder is
equivalent to the gas particles.

\begin{figure}[tbp]
\includegraphics[scale=0.5,angle=-90]{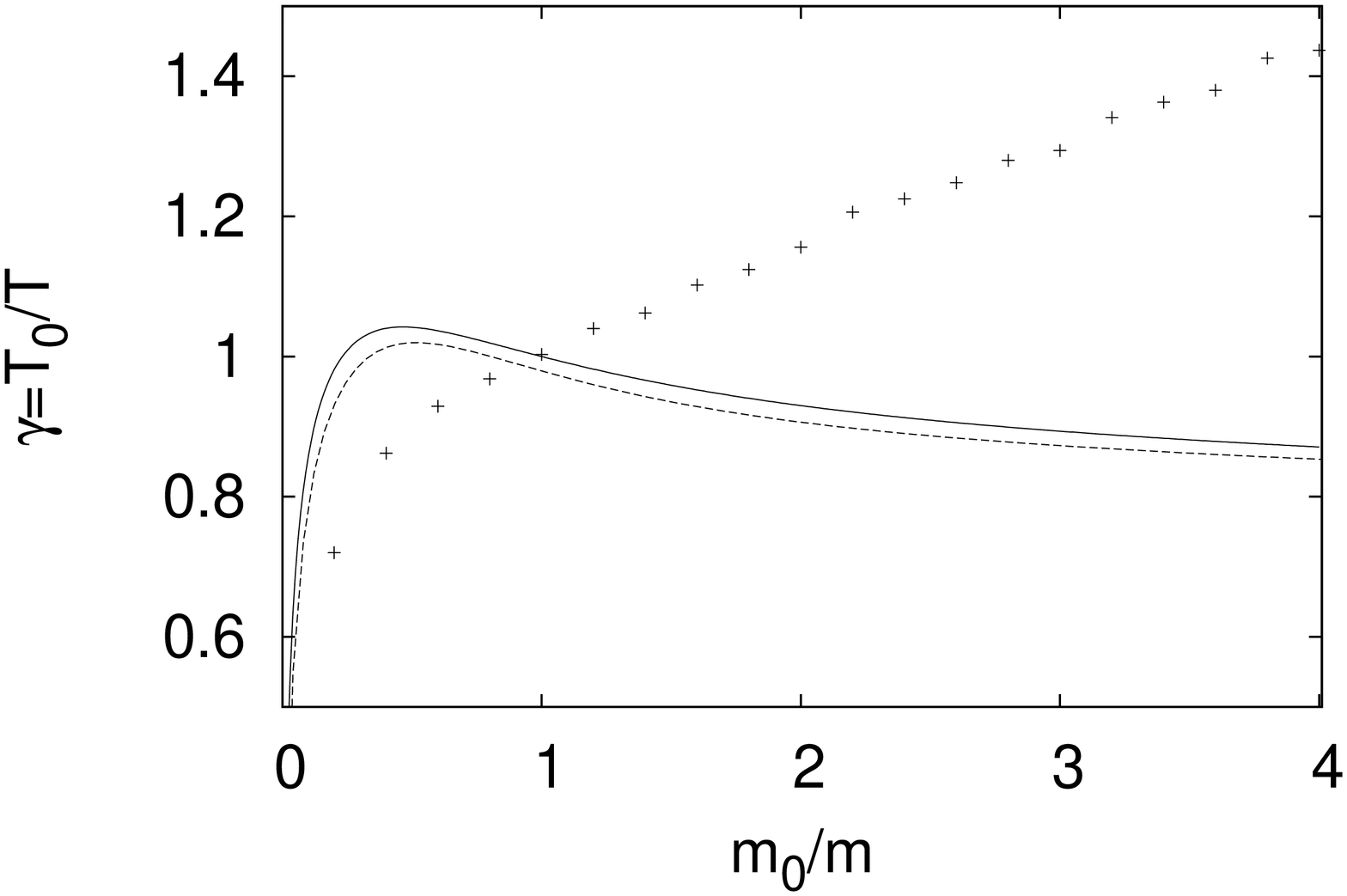}
\caption{Ratio between the impurity temperature $T_{0}$ and the host
gas temperature $T$ in the bulk of the system, once the Fourier
state has been reached, as a function of the ratio of the masses
$m_{0}/m$ . The coefficients of normal restitution for the gas-gas
and for the gas-intruder collisions are $\protect\alpha=
\protect\alpha_{0}=0.9$. The diameter of the intruder is the same as
the diameter of the host particles. The symbols are molecular
dynamics simulation results, the solid line the theoretical
prediction from the Boltzmann equation using a truncated Sonine
expansion, and the dashed line the weak inelasticity host gas limit
given in the Appendix and discussed in the text.} \label{fig8}
\end{figure}

\begin{figure}[tbp]
\includegraphics[scale=0.5,angle=-90]{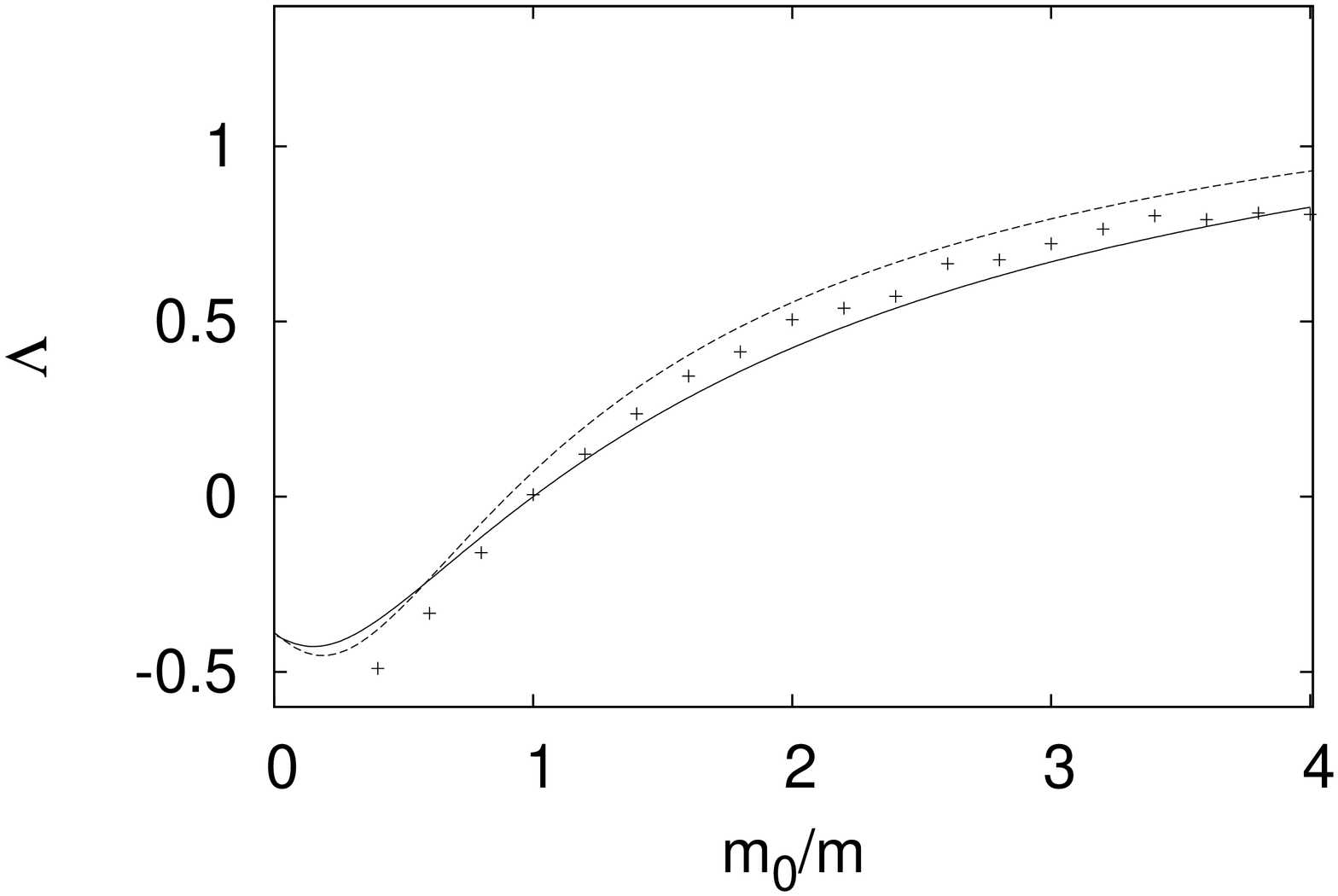}
\caption{The dimensionless thermal diffusion factor $\Lambda$ for
the same system as in Fig. {\protect\ref{fig8}}.} \label{fig9}
\end{figure}

The dependence on the mass ratio of both $\gamma$ and $\Lambda$ is
shown in Figs. \ref{fig8} and \ref{fig9}, respectively. Now it is
observed that while the theory accurately predicts the thermal
diffusion factor, it clearly
fails to describe the breakdown of the energy equipartition. When the mass $%
m_{0}$ of the intruder is larger that the mass $m$ of the host gas
particles, the simulation results indicate that $T_{0}/T$ grows
rather fast with $m_{0}/m$, while the theory predicts a weak
decrease remaining smaller than unity. Violations of energy
equipartition in homogeneous granular mixtures much stronger than
predicted by the existing kinetic theory models have been reported
lately \cite{ByR11}.

\begin{figure}[tbp]
\includegraphics[scale=0.5,angle=-90]{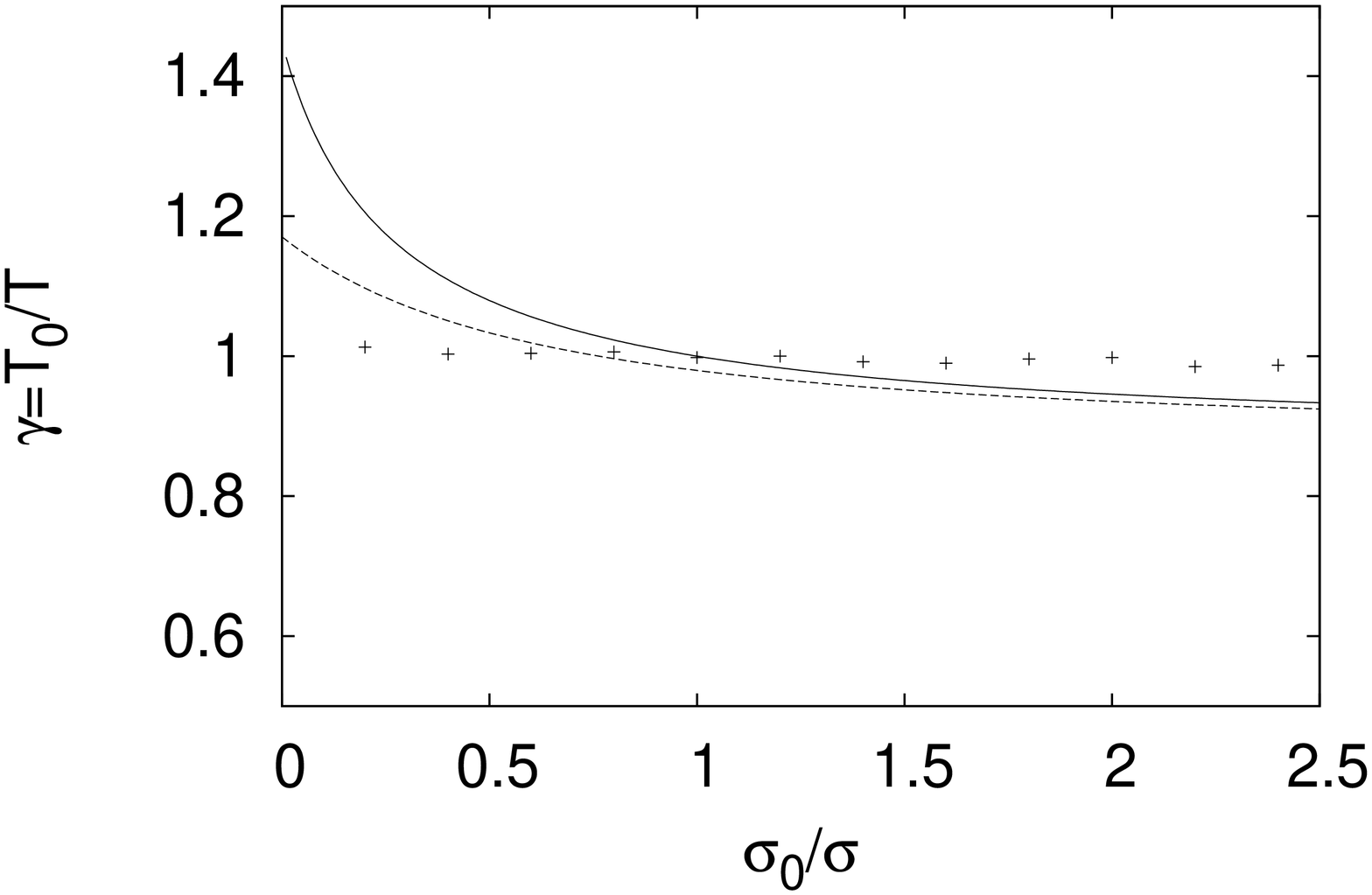}
\caption{Ratio between the impurity temperature $T_{0}$ and the host
gas temperature $T$ in the bulk of the system, once the Fourier
state has been
reached, as a function of the ratio of the diameters $\protect\sigma_{0}/%
\protect\sigma$ . The coefficients of normal restitution for the
gas-gas and for the gas-intruder collisions are $\protect\alpha=
\protect\alpha_{0}=0.9$. The mass of the intruder is the same as the
mass the host particles. The symbols are molecular dynamics
simulation results, the solid line the theoretical prediction from
the Boltzmann equation using a truncated Sonine expansion, and the
dashed line the weak inelasticity host gas limit given in the
Appendix and discussed in the text.} \label{fig10}
\end{figure}

\begin{figure}[tbp]
\includegraphics[scale=0.5,angle=-90]{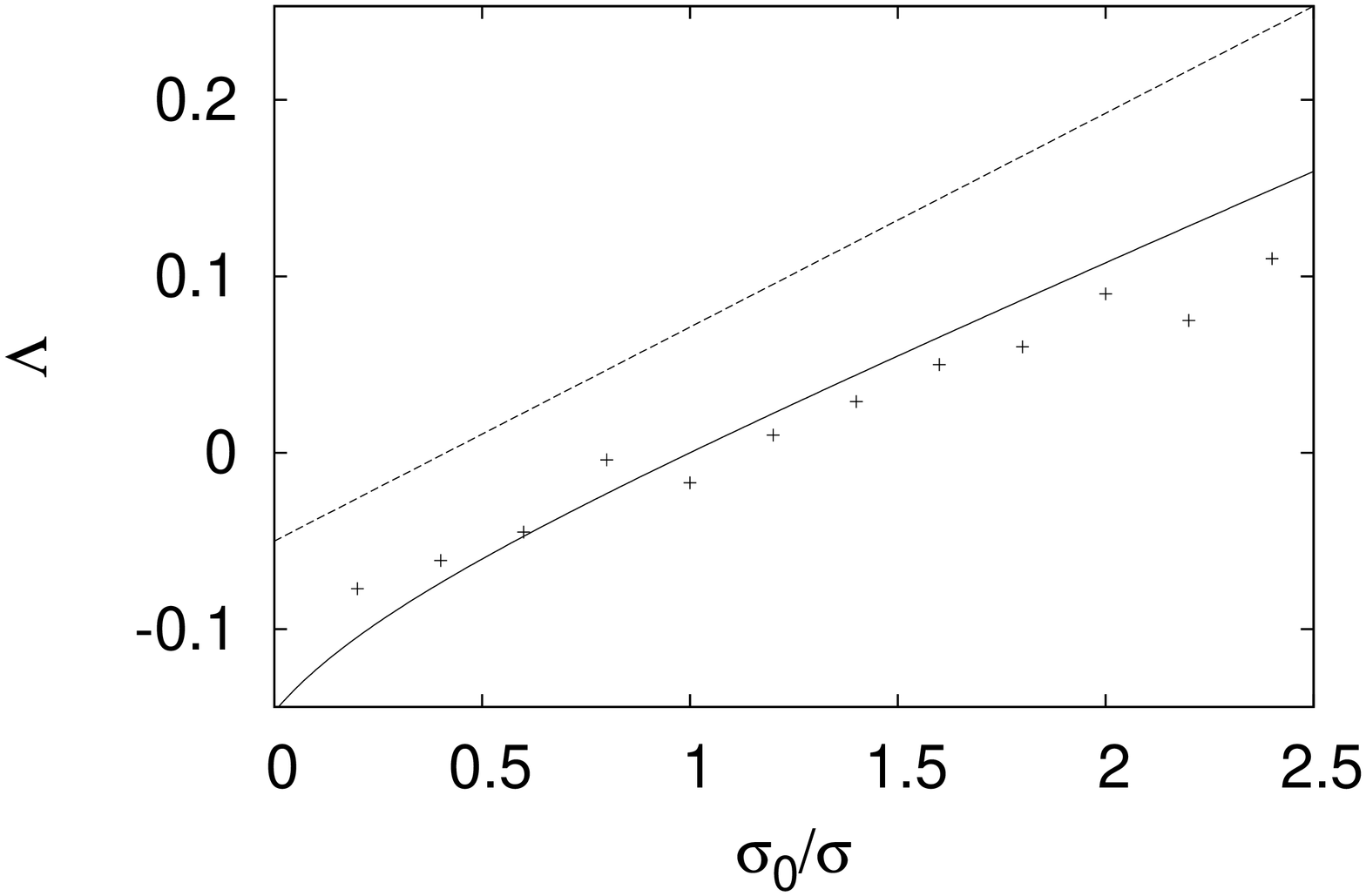}
\caption{The dimensionless thermal diffusion factor $\Lambda $ for
the same system as in Fig. {\protect\ref{fig8}}.} \label{fig11}
\end{figure}

Finally, the dependence of $\gamma $ and $\Lambda $ on the diameter ratio $%
\sigma _{0}/\sigma $ is given in Figs. \ref{fig10} and \ref{fig11},
respectively. Again, the theoretical prediction for $\Lambda $ can
be considered as quite satisfactory, but not that for the
temperature ratio.

\section{Discussion}

A theoretical description of thermal segregation has been explored
under very controlled conditions. The fundamental assumptions are
the validity of Boltzmann kinetic theory for a binary mixture of
inelastic, smooth, hard spheres or disks, and a special exact steady
"normal" solution. The kinetic theory is limited to low density
gases, which excludes many experimental conditions of interest.
However, it provides a useful testing ground for exploring the
rather large parameter space of binary mixtures. The normal solution
is restricted to conditions for which all space and time dependence
can be captured by the hydrodynamic fields, and hence excludes
boundary layers as discussed above. The only segregation mechanism
considered in the special solution here is a thermal gradient.
However, and important feature distinguishing the analysis from
others in the literature is the absence of any explicit restriction to
small temperature gradients, i.e. a description of thermal
segregation outside the Navier-Stokes limit. Still, the absence of
solutions for extreme dissipation (see Figs. \ref{fig1} and
\ref{fig2}) suggests an implicit limitation of the normal solution.

As expected, the MD simulation of a finite system with externally
imposed temperature gradient does not give the Fourier state
exactly, due to boundary layers. However, in the bulk of the system the Fourier
state is confirmed to good
accuracy. The comparison of MD and theoretical predictions  for the thermal diffusion factor $%
\Lambda $ shows good agreement across the parameter space of $\alpha
_{0},m_{0},\sigma _{0}$. In all cases conditions are found for
increased composition of the impurities both along and opposite the
temperature gradient. The corresponding comparison for the
temperature ratio $\gamma $ is less satisfactory, particularly for
variations of $m_{0}/m$ and $\sigma _{0}/\sigma .$ It might be taken
as a signature of the breakdown of the low order Sonine expansion
truncation, but then it is puzzling why this same approximation
should work so well for $\Lambda $. This discrepancy between theory
and simulation for $\gamma $ remains unexplained at this point.

\section{Acknowledgements}

This research was supported by the Ministerio de Educaci\'{o}n y
Ciencia (Spain) through Grant No. FIS2011-24460 (partially financed
by FEDER funds).

\appendix
\section{Nearly elastic host gas limit}
\label{ap1} The distribution functions of (\ref{2.2}) and
(\ref{2.5}) must be solutions to the Boltzmann and Boltzmann-Lorentz
equations, as described in \cite{BKyD11}. They are approximated by
the truncated Sonine expansions,
\begin{eqnarray}
\phi ({\bm c}) &\simeq &\pi ^{-d/2}e^{-c^{2}}\left[
1-a_{01}(c^{2}-dc_{x}^{2})+\left( \frac{d-1}{2}\,b_{01}+\frac{3}{2}%
\,b_{10}\right) c_{x}\right.   \nonumber \\
&&\left. -b_{01}c^{2}c_{x}-(b_{10}-b_{01})c_{x}^{3}\right] ,
\label{ap1.1}
\end{eqnarray}%
\begin{eqnarray}
\Phi ({\bm c}_{0}) &\simeq &\pi ^{-d/2}e^{-c_{0}^{2}}\left[
1-A_{01}(c_{0}^{2}-dc_{0x}^{2})+\left( \frac{d-1}{2}\,B_{01}+\frac{3}{2}%
\,B_{10}\right) c_{0x}\right.   \nonumber \\
&&\left. -B_{01}c_{0}^{2}c_{0x}-(B_{10}-B_{01})c_{0x}^{3}\right] ,
\label{ap1.2}
\end{eqnarray}%
with the constants appearing in these expressions determined from
moments of the kinetic equations. For $1-\alpha \ll 1$ a systematic
determination of
these coefficients has been carried out for $d=2$ in the Appendix of \cite%
{BKyD11}. The general results for arbitrary dimension are given
here:
\begin{equation}
a_{01}\sim A_{01}=\mathcal{O}(1-\alpha ),  \label{ap1.3}
\end{equation}%
\begin{equation}
b_{01}=b_{10}+\mathcal{O}(1-\alpha ),  \label{ap1.4}
\end{equation}%
\begin{equation}
b_{10}=\left[ \frac{2d(1-\alpha )}{d-1}\right] ^{1/2}+\mathcal{O}(1-\alpha ),  \label{ap1.5}
\end{equation}%
\begin{equation}
B_{10}=B_{01}+\mathcal{O}(1-\alpha ),  \label{ap1.6}
\end{equation}%
\begin{equation}
B_{01}=\frac{16(d-1)(\sigma /\overline{\sigma }%
)^{d-1}+27h(2-h)^{3/2}}{h^{1/2}\left[ 24(d+2)-8(d+8)h+27h^{2}\right] }%
\,b_{01}+\mathcal{O}(1-\alpha ).  \label{ap1.7}
\end{equation}%
All the dependence on the mass ratio $m/m_{0}$ and the coefficient
of restitution $\alpha _{0}$ occurs through the parameter
\begin{equation}
h\equiv \frac{m(1+\alpha _{0})}{m+m_{0}}.  \label{ap1.8}
\end{equation}%
Moreover, the coefficients $B$ and $C$ characterizing the
hydrodynamic profiles defined in Eqs. (\ref{2.3}) and (\ref{2.6})
have the expressions
\begin{equation}
B=\frac{2\pi ^{\frac{d-1}{2}}(d-1)}{\sqrt{2}\Gamma \left( \frac{d-1}{2}\right)
}\,b_{01}+\mathcal{O}(1-\alpha ),  \label{ap1.9}
\end{equation}%
\begin{eqnarray}
C &=&\frac{\pi ^{\frac{d-1}{2}}}{16\sqrt{2}\Gamma \left( \frac{d+4}{2}%
\right) }\,\left\{ h^{1/2}\left[ 48+4d(6-h)-(56-27h)h\right]
B_{01}\right.
\nonumber \\
&&-\left[ 48(\sigma /\overline{\sigma})^{d-1}(d-1)\right.   \nonumber \\
&&+\left. \left. (2-h)^{3/2}(8+4d+27h)\right] b_{01}\right\} +\mathcal{O}%
(1-\alpha ),  \label{ap1.10}
\end{eqnarray}%
while the temperature ratio reads
\begin{eqnarray}
\gamma  &=&\frac{1+\alpha _{0}-h}{2-h}\,\left\{ 1+\frac{B_{10}}{32d(2-h)}%
\left[ (48+4d(6-h)-(56-27h)h)B_{01}\right. \right.   \nonumber \\
&&\left. \left.
-\frac{(2-h)^{3/2}(8+4d+33h-6h^{2})}{h^{1/2}}\,b_{01}\right]
\right\} +\mathcal{O}(1-\alpha )^{3/2}.  \label{ap1.11}
\end{eqnarray}%
Notice that all the quantities above have been explicitly evaluated
up to
order $(1-\alpha )^{1/2}$, with the exception of the temperature ratio $%
\gamma $ that has been computed up to order $1-\alpha $. Finally,
the
thermal diffusion coefficient $\Lambda $ is obtained by substituting Eqs.\ (%
\ref{ap1.9}) and (\ref{ap1.10}) into Eq.\ (\ref{2.8}). This provides
an expression valid up to order $(1-\alpha )^{0}$.

\end{document}